\documentclass{cpbtex}

\usepackage{xcolor, soul}

\begin{document}

\title{Hidden symmetry operators for asymmetric generalised quantum Rabi models\thanks{Project supported by the Australian Research Council (Grant No. DP170104934 and DP180101040).}}

\author{Xilin Lu$^{1}$, Zi-Min Li$^{1}$, Vladimir V. Mangazeev$^{1}$,\\  
and Murray T. Batchelor$^{2,3}$\thanks{Corresponding author. E-mail:~murray.batchelor@anu.edu.au}\\
$^{1~}${Department of Theoretical Physics, Research School of Physics,}\\ {Australian National University, Canberra ACT 2601, Australia}\\  
$^{2~}${Mathematical Sciences Institute, Australian National University,}\\ {Canberra ACT 2601, Australia}\\ 
$^{3~}${Centre for Modern Physics, Chongqing University, Chongqing 40444, China}
}   


\date{\today}
\maketitle

\begin{abstract}
The hidden $\mathbb{Z}_2$ symmetry of the asymmetric quantum Rabi model (AQRM) has recently been revealed via 
a systematic construction of the underlying symmetry operator. Based on the AQRM result, we propose an ansatz for the 
general form of the symmetry operators for AQRM-related models. Applying this ansatz we obtain the symmetry operator 
for three models: the anisotropic AQRM, the asymmetric Rabi-Stark model (ARSM) and the anisotropic ARSM. 
\end{abstract}



\section{Introduction}

The quantum Rabi model (QRM) \cite{Xie2017,Braak2019},   
{describing a two-level atom interacting with a single mode bosonic light field, is 
central to a number of experimental platforms for the quantum simulation of light-matter interactions} \cite{Frisk2019,Forn2019,Blais2020}.
In the presence of a bias term, which breaks the parity symmetry induced level crossings of the QRM, 
the system is described by the asymmetric quantum Rabi model (AQRM). 
It has been observed \cite{Braak2011,Chen2012,Zhong2014,Maciejewski2014,Li2015,Wakayama2017} that level crossings  
reappear in the spectrum of the AQRM when the bias parameter $\epsilon$ takes special values, 
indicating the existence of a hidden symmetry of the AQRM. 
Similar hidden symmetry has been observed in other AQRM-related models \cite{Li2020}.
These generalised models include the asymmetric versions of the anisotropic QRM \cite{Tomka2014,Xie2014,Chen2021} (the anisotropic AQRM) 
and the Rabi-Stark model \cite{Eckle2017,Xie2019} (the ARSM), 
where the special values of the bias terms are given by conditions on the $\epsilon$ values, the so-called $\epsilon$-conditions \cite{Li2020}.
These observations indicate that the hidden symmetry in asymmetric light-matter interaction models 
 is no coincidence, rather it is a general phenomenon.
{By investigating tunnelling dynamics in the displaced oscillator basis, a strong connection has been found between the hidden symmetry and selective tunnelling 
in the AQRM and the ARSM} \cite{Li2020}.

A numerical study implied that any hidden symmetry operator of the AQRM must depend on the system parameters \cite{Ashhab2020}.
The way to construct the symmetry operator responsible for the hidden $\mathbb{Z}_2$ symmetry of the AQRM was found recently \cite{Mangazeev2020}. 
This approach provides a cornerstone for constructing symmetry operators in related models.
In this article, making use of an ansatz based on the AQRM results, 
we determine the symmetry operators for the other AQRM-related models  under their $\epsilon$-conditions \cite{Li2020}. 
We begin with a brief review of the AQRM results and propose the ansatz based on these results in Section 2. 
This ansatz is used to calculate the $\mathbb{Z}_2$ symmetry operators $J$ for the 
anisotropic AQRM and the ARSM in Sections 3 and 4, respectively. 
In Section 5, we discuss the combined anisotropic ARSM. 
Concluding remarks are given in Section 6.

\section{Hidden symmetry in the AQRM and general ansatz}

The AQRM is defined by the Hamiltonian 
\begin{equation}\label{AQRMHamiltonian}
   H_{\mathrm{AQRM}}= a^\dagger a+g \sigma_x(a^\dagger+a)+\Delta \sigma_z+\epsilon \sigma_x,
\end{equation}
where $g$ is the coupling strength, $2\Delta$ is the level splitting and $\epsilon$ is the bias field. 
The frequency of the light field is here scaled to unity ($\omega=1$).

When $\epsilon=0$, the model reduces to the QRM and possesses the ($\mathbb{Z}_2$) parity symmetry
\begin{equation}
   [P, H_{\mathrm{AQRM}}|_{\epsilon=0}]=0.
   \end{equation}
Here $P$ (which squares to the identity) is the parity operator
\begin{equation}
P=\sigma_z e^{ \mathrm{i} \pi a^\dagger a}.
\end{equation}
Depending on the corresponding eigenvalues of the parity operator, 
energy levels of the QRM separate into two distinct sectors, with crossings only between levels from different sectors.

The $\epsilon$-condition for the AQRM is when $\epsilon={M}/{2}$ for some integer $M$. 
At these special values, crossings between different energy levels are observed in the spectra.
The corresponding $J$ operators were found in \cite{Mangazeev2020} (see also \cite{RBW2021}). 
These symmetry operators take the form of $2\times 2$ matrices of polynomials in $a$ and $a^\dagger$, 
where the degree of the polynomials is equal to $M$. 
As examples, the case $M=0$ simply gives the parity operator $P$ and the case $M=1$ gives 
\begin{equation}
    J_{\epsilon=1/2}=\mathcal{P}\begin{pmatrix}
    a^\dagger-a+2g+\frac{\Delta}{g} & a^\dagger+a\\
    -a^\dagger-a & a-a^\dagger-2g+\frac{\Delta}{g}
    \end{pmatrix}.
\end{equation}
Here the operator $\mathcal{P}$ is defined as 
\begin{equation}
    \mathcal{P}=e^{\mathrm{i} \pi a^\dagger a}.
\end{equation}
This operator anti-commutes with bosonic operators $a^\dagger$ and $a$:
\begin{equation}
   a^\dagger \mathcal{P}=-\mathcal{P}a^\dagger, \qquad a \mathcal{P}=-\mathcal{P}a.
\end{equation}

One can check that the $J$ operators square to polynomials of $H_{\mathrm{AQRM}}$ of degree $M$, 
hence they generate $\mathbb{Z}_2$ symmetries. 
For example, for the $M=1$ case we have 
\begin{equation}
    J_{\epsilon=1/2}^2=4H_{\mathrm{AQRM}|_{\epsilon=1/2}}+4g^2+\frac{\Delta^2}{g^2}+2.
\end{equation}
Alternatively, we can divide the energy levels into two sectors based on the sign of the corresponding $J$-eigenvalue. 
By doing this, we see that only levels with eigenvalues of different signs cross, just like the QRM case.

Based on these results for the AQRM, we propose that $J$ operators for other AQRM-related models take a similar form. 
Since we are only discussing models with one atom, the explicit form of our ansatz is 
\begin{equation}\label{ansatz}
    J_M= \mathcal{P}\begin{pmatrix}
    \sum_{i,j}^M a_{i,j}(a^\dagger)^i a^j & \sum_{i,j}^M b_{i,j}(a^\dagger)^i a^j\\
    \sum_{i,j}^M c_{i,j}(a^\dagger)^i a^j & \sum_{i,j}^M d_{i,j}(a^\dagger)^i a^j
     \end{pmatrix},
 \end{equation}
where $a_{i,j},b_{i,j},c_{i,j},d_{i,j}$ are constants needing to be determined.

The process of calculating $J$ operators for a specific model is straightforward. 
First we need the $\epsilon$-condition for crossings to appear, 
this can be obtained by exploring the pole structure of Braak's $G$-function \cite{Braak2011} appearing in the analytic solution.
Then we solve the matrix equation 
\begin{equation}\label{MatrixEquation}
    [J_M,H]=0
\end{equation}
at different values of $\epsilon$ to determine the values for constants, giving the expression for the $J$ operator.

It needs to be emphasised that this method is on a case-by-case basis and that we cannot obtain general formulae for arbitrary $M$. 
In fact, it is still an open question, even for the AQRM case, if it is possible to find the general formula for the $J$ operator.

\section{Anisotropic AQRM}

We start with the anisotropic generalisation of the AQRM.
The anisotropic AQRM \cite{Li2020} allows for the tuning of the relative weight $\lambda$ between rotating and counter-rotating terms. 
An example application of such a tuning is the electronic spin-orbit interaction in semiconductors, 
where the Rashba and Dresselhaus spin-orbit interactions act like the different rotating terms under certain transformation.
 
The anisotropic AQRM Hamiltonian is 
 \begin{equation}
    H_{\mathrm{an}}=\begin{pmatrix}
    a^\dagger a+\Delta & g(\lambda a^\dagger+a)+\epsilon\\
    g(a^\dagger+\lambda a)+\epsilon & a^\dagger a-\Delta
    \end{pmatrix}.
\end{equation}
The AQRM is recovered by setting $\lambda=1$.

From \cite{Li2020}, the $\epsilon$-condition for this model is $\epsilon=\frac{M\sqrt{\lambda}}{1+\lambda}$ for some integer $M$. 
As mentioned in the previous section, we solve the matrix equation 
\begin{equation}
    [J_M,H_{\mathrm{an}}]=0
\end{equation}
to determine the constants $a_{i,j},b_{i,j},c_{i,j},d_{i,j}$ in the ansatz (\ref{ansatz}). 
In this way the $J$ operators for the two lowest orders are found to be
%
\begin{equation}
    J_1=\frac{1}{2(1+\lambda)}\mathcal{P}\begin{pmatrix}
    A_1^+ & B_1\\-B_1^\dagger & A_1^-
    \end{pmatrix},
\end{equation}
where the operators 
\begin{equation}
    \begin{split}
        A_1^\pm&=\pm 2g\sqrt{\lambda}(1+\lambda)(a^\dagger-a) \pm g^2(1+\lambda)^3+2\Delta(1+\lambda)+\lambda-1, \\
       B_1&= 2g(1+\lambda)(\lambda a^\dagger+a), 
    \end{split}
\end{equation}
and
\begin{equation} \label{J2}
    J_2=\frac{1}{4(1+\lambda)^2}\mathcal{P}\begin{pmatrix}
    A_2^+ & B_2\\B_2^\dagger & A_2^-
    \end{pmatrix},
\end{equation}
{where the operators $A_2^\pm$ and $B_2$ are given in Appendix A.}
The normalisation constants appearing in front of $\mathcal{P}$ 
ensure that these expressions simplify to the AQRM $J$ operators \cite{Mangazeev2020} when $\lambda=1$. 
It is tedious but possible to check that both of the above $J$ operators square to polynomials of $H$ of order $M$,  
hence we have found the $\mathbb{Z}_2$ hidden symmetry operators for the anisotropic AQRM.
{For example,}
\begin{equation}
J_1^2=g^2(1+\lambda)^2 H_{\mathrm{an}}+ \frac{g^2(1+\lambda)^2}{2}+\frac{g^4(1+\lambda)^4}{4}+\left(\Delta-\frac{1-\lambda}{2(1+\lambda)}\right)^2.
\end{equation}

We see that the expressions for the $J$ operators quickly become cumbersome as $M$ increases.
For this reason we only include the $M=1$ case for the other models covered in the following sections.

\section{Asymmetric Rabi-Stark model}

Another model that we discuss here is the ARSM, with Hamiltonian
\begin{equation}
 H_{\mathrm{ARSM}}=\begin{pmatrix}
 (1+U)a^\dagger a+\Delta & g(a^\dagger+a)+\epsilon\\
 g(a^\dagger+a)+\epsilon & (1-U)a^\dagger a-\Delta
 \end{pmatrix},
 \end{equation} 
where we have added the Stark term $U a^\dagger a \sigma_z$ to the AQRM Hamiltonian (\ref{AQRMHamiltonian}). 
Interestingly, this Stark term is adjustable in the Grimsmo-Parkins scheme \cite{Grimsmo2013} for the cavity quantum electrodynamics realisation.
Without the bias $\epsilon$, this model has $\mathbb{Z}_2$ symmetry and has been exactly solved \cite{Eckle2017,Xie2019}. 
Special behaviour such as selective interactions \cite{Cong2020} have also been explored. 
However, with the bias term the symmetry is again broken, which makes the analysis much harder, unless $\epsilon$ satisfies the $\epsilon$-condition.

Note that we need $|U|<\omega=1$ here to avoid unphysical results \cite{Xie2019}.
We therefore re-parametrise $U=\sin t$ to simplify later expressions.
Under this parametrisation, the $\epsilon$-condition \cite{Li2020} for this model is $\epsilon=\frac{M}{2}\cos t$. 
For the $M=1$ case, using the same method as above, we arrive at the symmetry operator  
\begin{equation}
    J_1=\mathcal{P} 
    \begin{pmatrix} A & B \\ C & D 
\end{pmatrix} ,
\end{equation}
where
\begin{equation}
    \begin{split} &A=\sin t(1+\sin t)a^\dagger a+g \cos t(a^\dagger-a)+(2g^2+\Delta(1+\sin t)),\\
    &B = g(a^\dagger+a)+\frac14 \,{\sin 2t}, \qquad C= -g(a+a^\dagger)+\frac14 \,{\sin 2t},\\
    &D = \sin t(1-\sin t)a^\dagger a-g \cos t(a^\dagger-a)-(2g^2-\Delta(1-\sin t)).
    \end{split}
\end{equation}
As expected, this result simplifies to the AQRM case when $t=0$.

The $J$ operators for the ARSM again square to polynomials of $H$. 
However, the interesting point here is that these $J$ operators have a relation of order $2M$ with the Hamiltonian instead of order $M$ for the AQRM case. 
For example,
\begin{equation}\label{2MRec1}
    J_1^2=\sin^2 t \, H_{\mathrm{ARSM}}^2 +(2\Delta\sin t+4g^2) H_{\mathrm{ARSM}}+C,
\end{equation}
where $C=4g^2\Delta\sin t+g^2\cos 2t+\Delta^2+4g^4+g^2$. 
The result for $J_2^2$ is given in Appendix A.
Since the coefficients of terms with orders higher than $M$ are functions of $t$, 
extra orders are considered as the result of the Stark term $U a^\dagger a \sigma_z$.

\section{Anisotropic asymmetric Rabi-Stark model}

Recently the anisotropic Rabi-Stark model is also drawing some attention \cite{Xie2020}. 
Knowing results for the special cases of the asymmetric version, 
it is worth establishing the symmetry operator for the anisotropic ARSM.
The Hamiltonian for this rather complicated model is 
\begin{equation}
H_{\mathrm{AARSM}}=\begin{pmatrix}
(1+\sin t)a^\dagger a+\Delta & g(\lambda a^\dagger+a)+\epsilon\\
g( a^\dagger+\lambda a)+\epsilon & (1-\sin t)a^\dagger a-\Delta 
\end{pmatrix},
\end{equation}
where we again use the parametrisation $U=\sin t$. 
Without much surprise, we observe that the $\epsilon$-condition is $\epsilon=\frac{M\sqrt{\lambda}}{1+\lambda}\cos t$, which is simply the combination of ingredients.

Following the same steps as before, we obtain
%
\begin{equation}\begin{split}
    &J_1=\mathcal{P}\begin{pmatrix}
    \sin t (1+\sin t)a^\dagger a+g \sqrt{\lambda}\cos t(a^\dagger-a)+\Bar{A}_+ & g(\lambda a^\dagger+ a)+\frac{\sqrt{\lambda}}{2(1+\lambda)}\sin 2t\\
    -g(a^\dagger+\lambda a)+\frac{\sqrt{\lambda}}{2(1+\lambda)}\sin 2t & \sin t (1-\sin t)a^\dagger a-g \sqrt{\lambda}\cos t(a^\dagger-a)+\Bar{A}_-
    \end{pmatrix},
\end{split}
\end{equation}
with constants
\begin{equation}
    \Bar{A}_\pm=\frac{1}{4}\left(\frac{\lambda-1}{1+\lambda}(1+\cos 2t) \pm 2g^2(1+\lambda)^2 +4\Delta(1\pm \sin t)\right).
\end{equation}

From the previous sections, we know that only the Stark term produces extra orders in the relation between $J$ and $H$. 
Therefore, the $J$ operators for the anisotropic ARSM follow similar $2M$-order relations with the Hamiltonian as the ARSM cases (\ref{2MRec1}) and (\ref{2MRec2}). 

\section{Conclusion and discussion}

In this letter we have demonstrated that the recent results for the AQRM $\mathbb{Z}_2$ symmetry operator  
can be generalised to other AQRM-related models, namely the anisotropic AQRM, 
the ARSM and the anisotropic ARSM. 
This confirms that the existence of hidden symmetry is a general phenomena, and not restricted to the AQRM. 
The method we have used is to assume that the underlying symmetry operators $J$ take similar forms as the AQRM case. 
The explicit form for this ansatz is given by (\ref{ansatz}). Starting with this ansatz, 
we have calculated the constants involved by solving the matrix equation (\ref{MatrixEquation}). 
This process is straightforward to perform but does not give a general expression. 
The procedure therefore needs to be repeated for each value of $M$ and for each model. 
We also show that $J_M^2$ is a polynomial in terms of the Hamiltonian, which defines the $\mathbb{Z}_2$ nature of the symmetry. 
Curiously, the degree of the polynomial can be affected by the terms added to the AQRM. 
Here we saw that the Stark term $U a^\dagger a \sigma_z$ changes the polynomial degree from $M$ to $2M$ 
while the anisotropic parameter $\lambda$ makes no change.

There are still many questions to explore. 
For example, how far can the AQRM be deformed while preserving the hidden symmetry, and what are the underlying mathematical structures \cite{RW2021}? 
In other work \cite{Lu2021}, we have studied the multi-qubit generalisation of the AQRM, also known as the biased Dicke model. 
We found that only the lowest non-trivial order ($M=1$) of the hidden symmetry is present, 
which we believe is due to the interference between atoms. 
In another direction, the multi-photon AQRM has also {recently} been investigated \cite{Xie2021}, where hidden symmetry with various orders {is} observed. 
Surprisingly, the spectral curve crossings appear in the three-dimensional $E,g,\epsilon$ plot instead of the usual $E,g$ plot. 
It will also be very interesting to look at the underlying symmetry operator structure of this model.
{Symmetry operators have now also been constructed within the Bogoliubov operator approach} \cite{XC2021}.
{We conclude by emphasizing that a precise physical interpretation of hidden symmetry operators remains to be determined, 
even for their simplest manifestation in the AQRM.}

\section*{{Appendix A: Collection of formulae}}

In this Appendix we collect some of the more lengthy results.
For the anisotropic AQRM, the operators appearing in (\ref{J2}) are 
\begin{equation}
    \begin{split}
        A_2^\pm&=\pm 4g^2(1-\lambda)^4a^\dagger a\pm 8g^2\lambda(1+\lambda)^2((a^\dagger)^2+a^2)\mp 4g\sqrt{\lambda}(1+\lambda)(g^2(1+\lambda)^3 \pm 2(\Delta \lambda+\lambda+\Delta-1)(-a^\dagger+a)\\
        &\phantom{=} ~ \pm (1+\lambda)\big(g^4(1+\lambda)^5+4\Delta(\Delta\lambda+\lambda+\Delta-1)+2g^2(1+\lambda)((1-\lambda)^2\pm2\Delta(1+\lambda)^2)\big),\\
     B_2&=8g^2(1-\lambda)\sqrt{\lambda}(1+\lambda)^2a^\dagger a+8g^2\lambda^{3/2}(1+\lambda)^2((a^\dagger)^2-a^2)+4g^3\lambda(1+\lambda)^4(a^\dagger+a)\\
     &\phantom{=} ~ +4\sqrt{\lambda}\big(g^2(1-\lambda)(1+\lambda)^2+2(\Delta\lambda+\lambda+\Delta-1)\big).
    \end{split}
\end{equation}
For the asymmetric Rabi-Stark model, the square of the $J_2$ operator is
\begin{equation} \label{2MRec2}
\begin{aligned}
    J^2_2=&\sin^4 t \cos^2 t H_{\mathrm{ARSM}}^4+\sin^2 2t(2g^2+\Delta \sin t)H_{\mathrm{ARSM}}^3+\cos^2 t \big(\sin^2 t(g^2+8g^4+6\Delta^2+\cos^4 t)\\
    &+g^2 \sin t(\sin 3t-4\Delta(\cos 2t-5)+16g^4)\big)H_{\mathrm{ARSM}}^2+2\cos^2 t\big(\Delta \sin t(\cos^4t+2\Delta^2)\\
    &+4g^4(6\Delta\sin t+\cos 2t+1)+g^2\Delta(\sin 3t+\sin t-4\Delta \cos 2t+8\Delta)+16g^6\big)H_{\mathrm{ARSM}}\\
    &+\cos^2t\big(\Delta^2\cos^4t+8g^6(4\Delta\sin t+\cos 2t+1)+2g^2\Delta^2(4\Delta\sin t+\cos 2t+1)\\
    &+4g^4\Delta(\sin 3t+\sin t-2\Delta\cos 2t+4\Delta)+\Delta^4+16g^8\big).
\end{aligned}
\end{equation}


\addcontentsline{toc}{chapter}{References}
\begin{thebibliography}{99}\footnotesize
\itemsep=-3pt plus.2pt minus.2pt   

\bibitem{Xie2017} Xie Q, Zhong H, Batchelor M T and Lee C 2017 
\href{https://doi.org/10.1088/1751-8121/aa5a65}{\emph{J. Phys. A: Math. Theor.} \textbf{50} 113001}

\bibitem{Braak2019} Braak D 2019 \href{https://doi.org/10.3390/sym11101259}{\emph{Symmetry} \textbf{11} 1259} 

\bibitem{Frisk2019} {Kockum A~F, Miranowicz A, Liberato S D, Savasta S and Nori F 2019} 
\href{https://doi.org/10.1038/s42254-018-0006-2}{\emph{Nat. Rev. Phys.} \textbf{1} 19}
	
\bibitem{Forn2019} {Forn-D\'iaz P, Lamata L, Rico E, Kono J and Solano E 2019} 
\href{https://doi.org/10.1103/RevModPhys.91.025005}{\emph{Rev. Mod. Phys.} \textbf{91} 025005}

\bibitem{Blais2020} {Blais A, Grimsmo A~L, Girvin S~M and Wallraff A 2021} 
 \href{https://doi.org/10.1103/RevModPhys.93.025005}{\emph{Rev. Mod. Phys.} \textbf{93} 025005}

\bibitem{Braak2011} Braak D 2011 
\href{https://doi.org/10.1103/PhysRevLett.107.100401}{\emph{Phys. Rev. Lett.} \textbf{107} 100401} 

\bibitem{Chen2012} {Chen Q-H, Wang C, He S, Liu T and Wang K-L 2012} 
\href{https://doi.org/10.1103/PhysRevA.86.023822}{\emph{Phys. Rev. A} \textbf{86} 023822} 

\bibitem{Zhong2014} Zhong H, Xie Q, Guan X, Batchelor M T, Gao K and Lee C 2014 
\href{https://doi.org/10.1088/1751-8113/47/4/045301}
{\emph{J. Phys. A: Math. Theor.} \textbf{47} 045301}

\bibitem{Maciejewski2014} {Maciejewski A~J, Przybylska M and Stachowiak T 2014} 
\href{https://doi.org/10.1016/j.physleta.2014.10.001}
{\emph{Phys. Lett. A} \textbf{378} 3445}

\bibitem{Li2015} Li Z-M and Batchelor M T 2015 
\href{https://doi.org/10.1088/1751-8113/48/45/454005}{\emph{J. Phys. A: Math. Theor.} \textbf{48} 454005}

\bibitem{Wakayama2017} Wakayama M 2017 
\href{https://doi.org/10.1088/1751-8121/aa649b}{\emph{J. Phys. A: Math. Theor.} \textbf{50} 174001}

\bibitem{Li2020} Li Z-M and Batchelor M T 2021 
\href{https://doi.org/10.1103/PhysRevA.103.023719}{\emph{Phys. Rev. A} \textbf{103} 023719}

\bibitem{Tomka2014} Tomka M, El Araby O, Pletyukhov M and Gritsev V 2014 
\href{https://doi.org/10.1103/PhysRevA.90.063839}{\emph{Phys. Rev. A} \textbf{90} 063839}

\bibitem{Xie2014} Xie Q-T, Cui S, Cao J-P, Amico L and Fan H 2014 
\href{https://doi.org/10.1103/PhysRevX.4.021046}{\emph{Phys. Rev. X} \textbf{4} 021046}

\bibitem{Chen2021} Chen X-Y, Duan L, Braak D and Chen Q-H 2021
\href{https://doi.org/10.1103/PhysRevA.103.043708}{\emph{Phys. Rev. A} \textbf{103} 043708}

\bibitem{Eckle2017} Eckle H-P and Johannesson H 2017 
\href{https://doi.org/10.1088/1751-8121/aa785a}{\emph{J. Phys. A: Math. Theor.} \textbf{50} 294004}

\bibitem{Xie2019} {Xie Y-F, Duan L and Chen Q-H 2019} 
\href{https://doi.org/10.1088/1751-8121/ab1cf6}{\emph{J. Phys. A: Math. Theor.} \textbf{52} 245304}

\bibitem{Ashhab2020} Ashhab S 2020 
\href{https://doi.org/10.1103/PhysRevA.101.023808}{\emph{Phys. Rev. A} \textbf{101} 023808} 

\bibitem{Mangazeev2020} Mangazeev V V, Batchelor M T and Bazhanov V V 2021 
\href{https://doi.org/10.1088/1751-8121/abe426}{\emph{J. Phys. A: Math. Theor.} \textbf{54} 12LT01}

\bibitem{RBW2021} Reyes-Bustos C, Braak D and Wakayama M 2021 
\href{https://doi.org/10.1088/1751-8121/ac0508}{\emph{J. Phys. A: Math. Theor.} \textbf{54} 285202}

\bibitem{Grimsmo2013} Grimsmo A L and Parkins S 2013 
\href{https://doi.org/10.1103/PhysRevA.87.033814}{\emph{Phys. Rev. A} \textbf{87} 033814}

\bibitem{Cong2020} Cong L, Felicetti S, Casanova J, Lamata L,  Solano E and Arrazola I 2020
\href{https://doi.org/10.1103/PhysRevA.101.032350}{\emph{Phys. Rev. A} \textbf{101} 032350}

\bibitem{Xie2020} Xie Y-F, Chen X-Y, Dong X-F and Chen Q-H 
\href{https://doi.org/10.1103/PhysRevA.101.053803}{\emph{Phys. Rev. A} \textbf{101} 053803}

\bibitem{RW2021} {Reyes-Bustos and Wakayama M}, 
 Degeneracy and hidden symmetry -- an asymmetric quantum Rabi model with an integer bias
 \href{https://arxiv.org/abs/2106.08916}{arXiv:2106.08916} 

\bibitem{Lu2021} Lu X, Li Z-M, Mangazeev V V and Batchelor M T 2021 
\href{https://doi.org/10.1088/1751-8121/ac0f16}{\emph{J. Phys. A: Math. Theor.} \textbf{54} 325202}

\bibitem{Xie2021} Xie Y-F and Chen Q-H 2021 
\href{https://doi.org/10.1103/PhysRevResearch.3.033057}{\emph{Phys. Rev. Research} \textbf{3} 033057}

\bibitem{XC2021} {Xie Y-F and Chen Q-H},  
General symmetry operators of the asymmetric quantum Rabi model within Bogoliubov operator approach,  
\href{https://arxiv.org/abs/2107.08937}{arXiv:2107.08937} 


\end{thebibliography}
\end{document}